# INDICATIONS OF DARK MATTER DERIVED FROM LARGE SCALE FLOWS

Marc Davis[1] and Adi Nusser[2]
[1] Depts. of Astronomy and Physics, U.C. Berkeley
marc@coma.berkeley.edu
[2] Institute of Astronomy, Cambridge
adi@mail.ast.cam.ac.uk

## ABSTRACT

Recent progress in the measurement of relative distances to galaxies has been quite substantial, and catalogs of 3000 galaxies with distances are soon to become available. The peculiar *velocity* field (deviations from Hubble flow) derivable from these catalogs, when compared to the peculiar *gravity* field derived from all sky redshift surveys of galaxies such as the 1.2Jy IRAS survey, leads to a unique and extremely powerful test of the density parameter $\beta \equiv \Omega^{0.6}/b_I$, where $b_I$ is the possible linear bias of the IRAS selected galaxies relative to the mass fluctuations. We review the status of these large scale flow measurements and present a new methodology to describe the two fields by means of an expansion in a set of orthogonalized functions describing a general potential flow to any chosen resolution. The parameters of the flow can be estimated by minimization of the $\chi^2$ describing the scatter of observed versus predicted linewidths from an inverse Tully-Fisher relation. By this method one can intercompare the gravity and velocity fields coefficient by coefficient, deriving a precise fit for the density parameter and an assessment of the degree of coherence between the fields. The present situation is transitory– different analyses of the same data are not yielding consistent results. Until this embarassment is untangled, estimates of $\beta$ should be taken with a large grain of salt.

## 1. INTRODUCTION

The study of large scale flows permits an estimate of the density parameter on a characteristic scale of $10 - 30 h^{-1}$ Mpc, which is thirty times larger than the central regions of rich clusters, where mass estimates can be derived on the basis of hydrostatic equilibrium or virial equilibrium. As is well known, mass estimates from cluster analysis imply a cosmological density of order $\Omega \approx 0.1 - 0.2$, assuming that galaxies in and out of clusters have similar mass to light ratios, or that the baryon fraction in the clusters is typical of the Universe as a whole (White *et al.* 1993). The large scale flow analysis would be expected to yield similar estimates of the density parameter if the dark matter fully congregates with the galaxies on cluster scales, but could well yield a different value if galaxy formation is more efficient in the denser regions or if the dark matter is too hot to cluster on any but the very largest scales.

Dekel (1994) and Strauss and Willick (1995) have recently reviewed this subject in detail, and we shall confine our remarks here to a few comments on the discrepancies of the $\beta$ derived from different analyses. A decade ago the quantitative analysis of large scale flows consisted largely of the question of Vir-gocentric flow (Davis and Peebles 1983), but now the focus has broadened to analysis of the details of the flow on a scale four times as large, which has led to new insights about large scale structure in the Universe. The observational situation is improving rapidly, and will continue to improve in the next few years, particularly if high precision distance estimators, such as SN-Ia (Riess *et al.* 1994), or surface-brightness fluctuations (Tonry *et al.* 1994) in early type galaxies, lead to useful catalogs. Giovanelli *et al.* (1994) are preparing a full sky sample of some 2400 galaxies with Tully-Fisher distances, while Willick *et al.* (1995) have compiled a sample of 2900 galaxies over the full sky (Faber *et al.* 1994). These large samples probe the velocity field to a redshift of approximately 8000 km/s, and the distance accuracy of the Tully-Fisher indicator is estimated to have a precision of 15-20%.

The cosmological interpretation of large scale flows is dependent on late time linear perturbation theory, which extensive testing has shown to be valid on scales $> 10h^{-1}$ Mpc. This is convenient, because linear theory is easy to work with and is the fundamental prediction of alternative models of large scale structure. If one confines the analysis to simple moments of the velocity field, such as a bulk flow or the alignment of the flow on a given scale with the dipole of the cosmic microwave background, the cosmological test is limited to questions of the likelihood that a given model would produce sufficient power to generate such a flow.

To sharpen the cosmological probe, it is instructive not simply to analyse the *power* in the velocity flow, but to compare the *alignment* of the velocity field with the inferred *gravitational* field derived from the observed galaxy distribution. Toward this end, complete redshift surveys of whole-sky galaxy catalogs are an essential ingredient. The IRAS selected 1.2Jy survey (Fisher *et al.* 1995) of 5300 galaxies covering 88% of the sky is the largest available catalog at present, but it should soon be superceded by at least two other catalogs (Santiago *et al.* 1994) (Saunders *et al.* 1995). The IRAS catalog is very suitable for this type of analysis because the 60 micron flux measured by IRAS is unaffected by galactic extinction and it is possible to select galaxies down to galactic latitude $|b| = 5°$ with only modest confusion. However, the IRAS sample is quite dilute, particularly in cluster centers, and the inferred gravitational field has considerable noise, especially for structures with redshift $cz > 6000$ km/s.

If one had full knowledge of the mass fluctuation field $\delta_\rho(\mathbf{r})$ over all space, one could trivially write the gravitational fluctuation field $\mathbf{g}(\mathbf{r})$ as

$$\mathbf{g}(\mathbf{r}) = G\bar{\rho} \int d^3\mathbf{r}' \delta_\rho(\mathbf{r}') \frac{\mathbf{r}' - \mathbf{r}}{|\mathbf{r}' - \mathbf{r}|^3} \; , \qquad (1)$$

where $\bar{\rho}$ is the mean mass density of the Universe. The great simplification of late time linear theory is the intuitive relation between the gravity field $\mathbf{g}$ and the velocity field $\mathbf{v_L}$, namely the equating of the velocity as the acceleration times the time, where the only possible time is the Hubble time. The correct expression (Peebles 1980) is

$$\mathbf{v_L}(\mathbf{r}) = \frac{2}{3H_0 \Omega^{0.4}} \mathbf{g}(\mathbf{r}) \; . \qquad (2)$$

Assuming the galaxy distribution at least approximately traces the mass on large scale, with linear bias $b$ between the galaxy fluctuations $\delta_G$ and the mass fluctuations, ($b = \delta_G/\delta_\rho$), and replacing the integral over space with a sum over the galaxies in a catalog, with radial selection function $\phi(r)$, we have

$$\mathbf{v_L}(\mathbf{r}) = \frac{H_0}{4\pi\bar{n}} \sum_i \frac{1}{\phi(r_i)} \frac{\mathbf{r_i} - \mathbf{r}}{|\mathbf{r_i} - \mathbf{r}|^3} \; , \qquad (3)$$

where $\bar{n}$ is the mean galaxy density in the sample. Note that the result is insensitive to the value of $H_0$, as the right hand side has units of velocity. Thus in this game it is traditional to quote distances in units of km/s.

Because the effective weight per galaxy $1/\phi(r)$ diverges at large distance, one must cutoff the calculation at some limiting redshift, which for the 1.2Jy IRAS sample is usually chosen to be 20,000 km/s. To limit the direct influence of one galaxy on another and to filter out nonlinear effects, it is necessary to smooth the small scale interaction, typically with a smoothing scale of $r_s \geq 500$ km/s. The method assumes that the associated mass per galaxy is constant, but in fact it is straightforward to piece together different redshift catalogs with different weights and selection functions.

Two different schemes have been used to compute $\mathbf{v_L}$ from redshift catalogs. The original scheme of Yahil *et al.* (1991) solved Equation (3) by iteration, adiabatically turning on the gravity field to advance the position of the galaxies from redshift space to real space. Recent refinements on this method by computation on a grid (Strauss and Yahil 1994) will allow nonlinear corrections to be applied to the predicted velocity field. An alternative formulation that can be applied directly in redshift space has recently been suggested by Nusser and Davis (1994a). This method solves a modified form of the Poisson equation using spherical harmonic decomposition and can readily deal with variable smoothing. To lowest order the two methods agree, but differ on how they treat the triple valued zones, regions around cluster centers in which three separate distances can have the same redshift. The Nusser and Davis algorithm intrinsically assumes a monotonic relation between distance and redshift, thereby eliminating multivalued zones. The older, iterative algorithm allows the existence of multivalued zones but has no idea of how to position an object that finds itself in such a zone; several ad hoc algorithms have been adopted over the years. There is no good solution for this problem which is one reason to filter out nonlinear effects or to compare alternative algorithms. Fortunately, most of the local density field appears to be single valued when smoothed on scales of 500 km/s or larger; only around cluster centers is the reconstruction of questionable validity. Nonlinear reconstruction algorithms are of marginal utility because they do not eliminate the problems of the multi-valued zones, yet the nonlinear effects are only important in regions which are nearly identical to the multivalued zones. Furthermore these are the regions where the assumption of linear bias is most questionable.

## 2. $\delta - \delta$ COMPARISON

The POTENT algorithm (Dekel *et al.* 1990) is a scheme to convert the observed radial velocity field into a density field, and depends on the reasonable assumptions that the large scale velocity field is a potential flow in the linear limit. The assumption of potential flow allows one to construct a three dimensional velocity field from the observed radial flow, and the divergence of this field yields an estimate of the fluctuating *mass* field driving the flows. Comparison of this mass field with the observed galaxy distribution leads to an estimate of $\beta$. Details are given by Dekel (1994); the method has been applied to the Mark-2 catalog (Faber and Burstein 1988) by Bertschinger *et al.* (1990) and then compared to the galaxy density field derived from the 1.9Jy IRAS redshift catalog (Strauss *et al.* 1992), resulting in a very high density estimate, $\beta = 1.3 \pm 0.3$ (Dekel *et al.* 1993).

In order to define the velocity potential, it is necessary to heavily smooth the velocity data; Dekel *et al.* (1993) use a guassian smoothing of $\sigma = 1200$ km/s, which is sufficiently large to filter out the infall of the local group toward the Virgo supercluster! In its present implementation the POTENT method depends on a forward Tully-Fisher approach, in which a calibrated Tully-Fisher relation is used to infer the absolute magnitude of a galaxy given an observed line width. As is well known, this procedure suffers from homogeneous, and more particularly, inhomogeneous Malmquist bias (Schechter 1980). If the scatter in the distance estimator is accurately known, the Malmquist biases can be removed. If the incorrect scatter is used, of if no correction is applied, an overdensity of points in redshift space will appear to have the signature of a massive cluster, even if the cluster has no mass at all. Dekel *et al.* (1993) undertook an extensive series of Monte-Carlo tests to calibrate these biases, and their result is consistent with a number of earlier comparisons of IRAS predictions to peculiar velocity data (e.g. Kaiser *et al.* 1991, Straus 1988, Strauss and Davis 1988).

## 3. v-v COMPARISONS

Because this result is the only observational evidence for a high density Universe, it is certainly worth pursuing alternative methods that could provide confirming or refuting evidence. In the interval since Dekel *et al.* (1993), the available database has improved considerably. The IRAS sample has doubled with the availability of the 1.2Jy survey (Fisher 1992), but more importantly, the Mark-2 sample of 496 galaxies is being superceded by the Mark-3 sample of 2900 galaxies (Faber *et al.* 1994, Willick etal 1995a). Several new analyses are underway and results will be reported in due course. Here we give a progress report.

Willick *et al.* (1995b) compare the observed distribution of the magnitudes, redshifts, and linewidths for the Mark-3 data, and have cast the problem in terms of a likelihood of measuring a set of observed magnitudes, given the positions of the galaxies in redshift space and observed linewidths. The method accounts for the inhomogeneous space distribution by means of the observed low resolution density field measured from the IRAS sample. They have devised an algorithm that explicitly attempts to neutralize the effects of the triple valued zones in a statistical sense. Distortions from real to redshift space are provided by flow models derived from the IRAS gravity field, with $\beta$ the only adjustable parameter. The procedure then asks which value of $\beta$ is most likely to fit the observed distribution. The test favors a tentative value of $\beta = 0.55 \pm 0.13$. Details are given by Strauss and Willick (1995). While the method provides a powerful measure of which IRAS flow model is the best fit to the data, it does not in itself provide an independent picture of the flow field, which POTENT does provide, and which gives POTENT such a visual appeal.

Two alternative schemes of intercomparing the velocity and gravity fields have recently been presented by Nusser and Davis (1994a,b). The first is limited to a consideration of the dipole component of the radial peculiar velocity field and depends on a simple, but little used fact of potential theory. Consider a decomposition of the radial field on a given redshift shell into spherical harmonics. Let the shell have radius $r$ and position a charge (mass) at some location $\mathbf{R}$. For a potential field $\Phi$ satisfying the Laplace equation, the $l^{th}$ multipole $\Phi_l$ will scale as $r^l$ or $r^{-(l+1)}$, depending on whether $R > r$ or $r < R$, and the gradient of the potential ($\mathbf{v}$ or $\mathbf{g}$) will scale as $r^{l-1}$ or as $r^{-(l+2)}$. We are used to the iron sphere theorem of Newton: for $l = 0$, a particle feels no force from the mass external to it. Note that for $l = 1$, $d\Phi_1/dr$ is constant for $r < R$. That is, by transforming to the local group frame, the dipole amplitude of the velocity on all shells with $r < R$ is zero, and the influence of an external charge is measurable only for $l \geq 2$. In the local frame, the measured dipole on a given shell must be due entirely to the distribution of charge (mass) internal to that shell.

Such a situation overcomes a serious objection to the use of Equation (3)-that the gravity field is nonlocal. Since the estimates derived from IRAS do not integrate the mass density over all space, they might be missing some large scale components. This problem is explicitly avoided by limiting the consideration to the behavior of the dipole amplitude. Nusser and Davis (1994a) extracted the dipole component of the velocity field from an early version of the Mark-3 POTENT map and compared it to the dipole field of the IRAS gravity maps, for various values of $\beta$ (Figure 1). They find that the velocity and dipole vector directions are aligned within $11°$, and furthermore that wiggles in the amplitude of the dipole versus redshift $z$ are remarkably consistent. Note that the POTENT dipole does not go to zero at the origin because of the large smoothing. $\beta = 0.6 \pm 0.2$ seems to be the preferred fit, in agreement with Willick *et al.* (1995). One very important point from Figure 1 is that 450 km/s of shear is clearly generated within a distance of 5000 km/s, a distance to which the IRAS density field should be reliable. Only the 250 km/s difference of this measured shear amplitude and the 620 km/s dipole amplitude of the CMBR dipole can be induced by larger scale bulk flows. This must be kept in mind in the interpretation of claims of the detection of a bulk flow on a much larger scale (e.g. Lauer & Postman 1994).

## 4. INVERSE TULLY-FISHER RECONSTRUCTION OF A SMOOTHED FIELD

To generalize this approach, one might consider directly comparing the velocity field derived from POTENT with the IRAS gravity field for higher multipoles. This approach is not advisable, because the distribution of points comprising the Mark-3 sample are not uniformly distributed on the sky or in redshift, so that the multipoles are not orthogonal. Furthermore POTENT is constructed on a cubic grid, which is not suitable for higher order multipole comparison.

Consider an alternative approach that starts afresh with the measured magnitudes and linewidths. Nusser and Davis (1994b) describe an inverse Tully-Fisher algorithm that based on a flow model which is a general, smoothed description of the velocity field in redshift space. Assume that the Tully Fisher relationship is linear, i.e.

$$\eta_i \equiv ln(\Delta v_i) = \eta_0 + sM_i \pm \epsilon ,\qquad(4)$$

where $\eta_0$ and $s$ are the unknown zero point and slope of the regression, $\epsilon$ is the error in the relation with $\langle \epsilon \rangle = 0$, and

$$M_i = m_i - 5log(z_i - u_i) - 15 .\qquad(5)$$

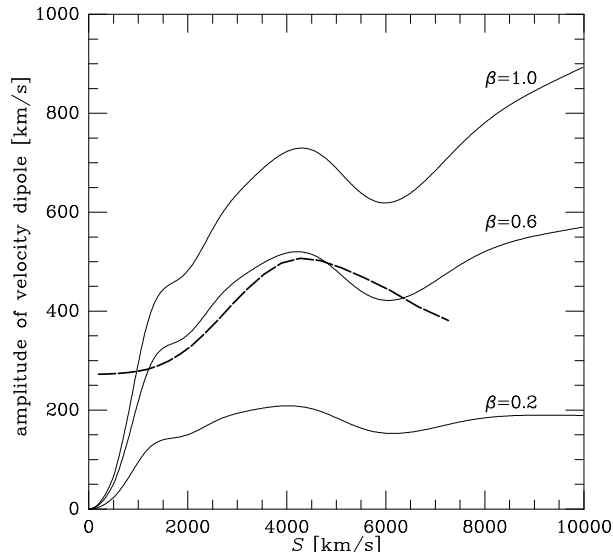

Fig. 1: The dipole amplitude of the Mark-3 sample compared to the IRAS predicted dipole amplitude for various $\beta$ values.

Here $m_i$ is the apparent luminosity, $z_i$ is the redshift, and $u_i$ is the peculiar velocity of the $i^{th}$ object in the LG frame. Let $M_i = M_{0i} + P_i$ with $M_{i0} = m_i - 5 log(z_i) - 15$ and $P_i = -5 log(1 - u_i/z_i)$. Then, if the TF scatter of galaxies in the $\eta$ direction is $\sigma_\eta$, one can compute a $\chi^2$ statistic that minimizes the $\eta$ scatter as a function of the parameters that describe the peculiar velocity field $P$. Such an *inverse* fit to the Tully-Fisher relationship does not suffer the usual Malmquist biases, but is biased in the presence of a "hot" velocity field or if there is a selection in the sample based on $\eta$. The typical bias in the inferred velocity is $\Delta u_{bias} \approx \sigma_u^2/z$. Given that the rms scatter of distances at a given redshift $\sigma_u$ is thought to be less than 300 km/s in most regions of redshift space, the bias of this technique is quite modest.

The problem with inverse methods is that, in contrast to a forward method such as POTENT, one does not generate pictures, but can only fit parameters to a model. However, with large datasets such as Mark-3, which have nearly full sky coverage to substantial depths, this is not a serious limitation. Consider for example an expansion of $P$ in terms of a set of functions that are orthogonal over a given dataset,

$$P_i = \sum_j \alpha_j \tilde{F}_i^j$$

where the $\alpha_j$ are the coefficients and $\tilde{F}_i^j$ is the value of the $j^{th}$ orthogonal function for object $i$, and

$$\sum_i \tilde{F}_i^j \tilde{F}_i^{j'} = \delta_K^{jj'} \;,$$

with $\delta_K$ the Kronecker delta function. Orthogonal functions are a convenient choice, as the $\chi^2$ minimization is linear in all the coefficients $\alpha_j$ as well as $\eta_0$ and $s$, and each term of the orthogonal expansion is decoupled from the others so that terms can be added one by one.

A natural choice of functions for a general expansion of the radial peculiar velocity field is again motivated by potential theory. Because we are working in the local group frame, $P(0) = 0$. One can guarantee the correct asymptotic behavior near the origin by expanding the field $P$ in terms of the derivatives of spherical Bessel functions times spherical harmonic functions $Y_{lm}(\theta, \phi)$,

$$P(z, \theta, \phi) = \sum_{nlm} a_{nlm} F_i^{nlm} = \sum_{nlm} a_{nlm} \frac{j'_l(k_n z)}{z} Y_{lm}(\theta, \phi) . \qquad (6)$$

See ND2 for details. One can choose $k_n$ by the desired boundary conditions at some $cz_{max} = 10000$ km/s, such as requiring $F(z_{max}) = 0$, but the solutions are insensitive to the details of this choice since the Mark-3 sample sets little constraint on the flow at this redshift. Fisher *et al.* (1994) have used similar sets of wavefunctions to describe the density field of the IRAS 1.2 Jy catalog. The choice of radial wave function can be flexible, as the method is only meant to be a fitting formula, and the irregular distribution of galaxies in redshift space prevents the original basis functions from being orthonormal. A convenient choice that leads to lower $\chi^2$ with fewer degrees of freedom is to use radial wavefunctions $j'(k_n y)/z$ where $y = (\ln(1 + (z/2000)^2))^{1/2}$. This has the effect of stretching the oscillations of the radial wavefunctions at larger redshift, allowing more radial resolution in the foreground than in the background, so as to match the resolution gradient of the gravity field and the absolute accuracy gradient of the observed peculiar velocities. By working to order $n_{max} = 5$, $l_{max} = 4$, we have 125 independent modes. Five additional modes describing an external quadrupole field ($F = $ constant $\times Y_{2m}(\theta, \phi)$) are also included for good measure. We transform the functions $F_i^j$ into an orthogonal basis set $\tilde{F}_i^j$ by means of an SVD inversion (Press *et al.* 1992).

To confirm that the method works, we constructed 2600-point mock Mark-3 catalogs derived from nbody simulations, including inhomogeneous sky selection and scatter in the TF relation of $\sigma_\eta = 0.03$. Of the 130 modes, only approximately 80 are determined with significance greater than one sigma, but these 80 coefficients contain a full description of the velocity field to the chosen resolution. The scatter in $\eta$ about the fit solution is spatially random in the simulations, indicating that fits to this resolution are adequate for the size and precision of the catalog. More detailed fits can be performed with larger set of Tully-Fisher data, or with more precise distance indicators.

To test whether the estimated mode amplitudes are correct, we have evaluated each mode amplitude using the true peculiar velocity of each simulation point. Figure 2a shows the comparison of the true versus ITF inferred mode amplitude for the 130 coefficients derived from the mock catalog. No bias in the mode amplitudes are detected, and the scatter of the amplitudes is completely consistent with the expected noise. Furthermore, most of the power of the large scale flow is contained within a few basis functions. For the true sky, we of course do not know a priori the true peculiar velocity of each point, and we can only substitute the linear-theory predicted peculiar velocity derived from the IRAS gravity maps. Figure 2b shows this test with the same mock Mark3 catalog, now substituting an IRAS derived estimate of the peculiar velocity of each point in the evaluation of the mode amplitudes. A flux limited mock IRAS catalog with the same selection function as the 1.2Jy survey was extracted from the simulation data and used to estimate the gravity field, so this test is a complete simulation of the real situation. Now a comparison of ITF inferred versus IRAS inferred mode amplitude shows more scatter, but still a complete absence of bias. The simulation is evaluated for $\beta = 1$, the true answer in this case. Evaluating the gravity field for other $\beta$ values changes the slope of the scatter diagram, so estimation of $\beta$ is completely straightforward. A gravity field evaluated using $\beta = 0.6$ is shown as the open symbols, and is clearly a poor fit to the diagonal line. With a suitable catalog of distance measurements, this method should lead to a definitive measure of $\beta$.

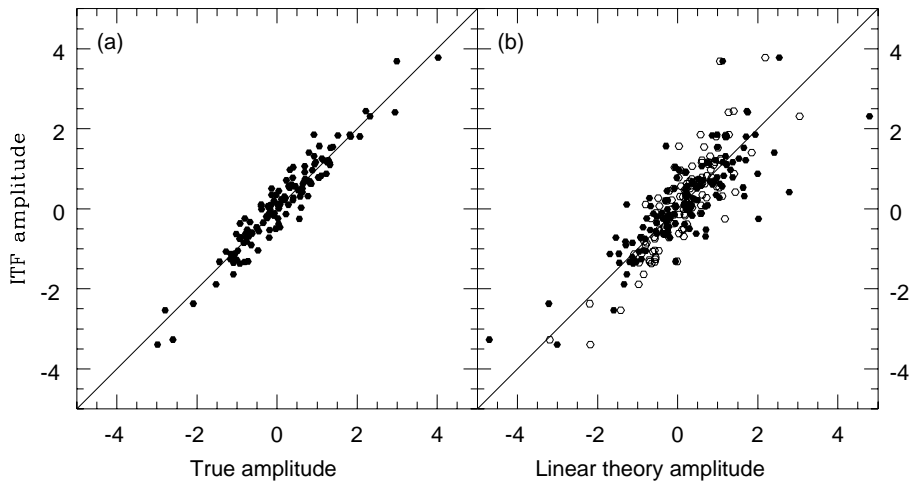

Fig. 2: (a) The ITF inferred amplitude of the orthogonal modes of a mock Mark-3 catalog are plotted versus the true mode amplitude. (b) The solid symbols show the ITF inferred amplitudes versus the amplitude measured from the linear theory predictions for $\beta = 1$ (the correct answer in this simulation). The scatter is larger than Fig. 2a, but the slope remains unbiased. The open symbols are for a gravity field with $\beta = 0.6$; note that the diagonal line is now a poor fit.

The great advantage of this scheme is that it allows the comparison of the velocity and gravity field mode by mode, with identical resolution to each field. The Monte-Carlo simulations with realistic mock catalogs confirm that the bias in the extracted velocity field is negligible. The method allows for the gradient in radial resolution of the two fields, so it permits the extraction of the maximum useful information for each field. One can rank order the modes by their signficance, or contribution to $\chi^2$, keeping only the most significant modes, so as to provide a very compact description of the field. This is an example of a Karhunen-Lo'eve transformation (Therrien 1992). Recent applications of this technique to astrophysics include the CMBR fluctuations of the COBE data (Gorski 1994, Bunn 1994, Bond 1994), the power spectral analysis of fluctuations in redshift surveys (Vogeley 1994), and the spectral classification of galaxies (Connolly *et al.* 1994).

However, at the present time, this method is not yielding sensible results when applied to the Mark-3 catalog. There is strong coherence in the residuals between the ITF velocity field and the gravity field which is unphysical and suggestive of a calibration error. Work is in progress to iron out the problem and results will hopefully be available soon.

## 5. WHY DON'T THE ALTERNATIVE METHODS YIELD A CONSISTENT $\beta$ ?

The POTENT-IRAS density comparison resulted in an estimate of $\beta > 1$, while the v-v comparisons discussed above are giving a best value half as large. What could account for such a discrepancy?

It is important to note that, while the different methods might start with the same basic data, their complex processing is rather different. In the $\delta - \delta$ analyses, the weighting is dominated by infall to cluster centers and outflow from voids. Larger scale components of the velocity field are less important than the local divergence. The direct velocity-velocity comparisons, on the other hand, are dominated by the signature of the dipole reflex of the motion of the local group. Figure 1 shows that 2/3 of the motion of the LG is reflected within the galaxy distribution to redshift 6000 km/s. Giovanelli *et al.* (1994) are claiming the reflex dipole is larger than shown in Figure 1, which would increase the inferred $\beta$ toward the POTENT value. The coherent residuals between the ITF inferred field and the IRAS predicted field for the Mark-3 sample are certainly cause for concern. Until we understand better the reason for this discrepancy, we must beware of possible systematic errors in $\beta$ values so derived.

The subject of large scale flows is rapidly evolving. The large datasets from Giovanelli *et al.* (1994) and Mark-3 are likely to be released soon, and other, more accurate catalogs of peculiar velocities (e.g. using surface-brightness fluctuations) are under construction. The 1.2Jy IRAS catalog will be soon superceded by the IRAS catalog of 15,000 galaxies with flux greater than 0.6Jy (Saunders *et al.* 1994) and by the ORS/IRAS catalog (Santiago *et al.* 1994), both of which will offer denser sampling and therefore improved statistical precision to the gravity field. The large scale flow analyses should yield a $\beta$ value accurate to 10%, and the various estimates should certainly be consistent. Since progress continues at a rapid pace, it is perhaps premature to expect consistency on this important parameter at the present moment.

## ACKNOWLEDGEMENTS

This research was supported by grants from the NSF and NASA.

## REFERENCES

Bertschinger, E., Dekel, A., Faber, S., Dressler, A., & Burstein, D 1990, ApJ 364, 370
Bond, J. R. 1994, Les Houches Lectures, preprint
Burstein, D. 1990, Rep. Prog. Phys., 53, 421
Bunn, T., 1995, Ph.D thesis, UC Berkeley
Connolly, A. J., Szalay, A. S., & Bershady, M. A. 1995, AJ, in press
Davis, M. & Peebles, P. J. E. 1983, Ann. Rev. of A&A, 21, 109
Dekel, A., Bertschinger, E., & Faber, S. 1990, 364, 349
Dekel, A. 1994, Ann. Rev of A&A, 32, 371
Dekel, A., Bertschinger, E., Strauss, M., Yahil, A., Davis, M., & Huchra, J. 1993, ApJ, 412, 1

Faber, S.M., & Burstein, D. 1988, in *Large Scale Motions in the Universe: A Vatican Study Week*, eds. V. C. Rubin & G. V. Coyne, (Princeton University Press), 116
Faber, S.M., Courteau, S, Dekel, A., Dressler, A., Kolatt, T., Willick, J. & Yahil, A 1994, in *Cosmic Velocity Fields*, eds. F. Bouchet & M. Lachi'eze-Rey, (Gif-sur-Yvette Cedex; Editions Frontieres), 15
Fisher, K. 1992, Ph.D thesis, UC Berkeley
Fisher, K., Scharf, C., & Lahav, O. 1994, MNRAS, 266, 219
Fisher, K., Huchra, J., Strauss, M., Davis, M., Yahil, A., & Schlegel, D. 1995, ApJ Supp, submitted
Giovanelli, R., Haynes, M., Salzer, J., Wegner, G., da Costa, L., & Freudling, W. 1994, AJ, 107, 2036
Gorski, K. 1994, ApJL. , 430, L85
Kaiser, N., Efstathiou, G., Ellis, R., Frenk, C., Lawrence, A., Rowan-Robinson, M., & Saunders, W. 1991, MNRAS, 252, 1
Lauer, T., & Postman, M. 1994, ApJ, 425, 418
Nusser, A., & Davis, M. 1994a, ApJ Lett, 421, L1
Nusser, A., & Davis, M. 1994b, MNRAS, submitted
Press, W., Teukolsky, S., Flannery, B., & Vetterling, W. 1992, *Numerical Recipes, The Art of Scientific Computing*, 2nd edition, Cambridge University Press
Riess, A. G., Press, W., & Kirshner, R. 1994, ApJ Lett., submitted
Santiago, B. X., Strauss, M., Lahav, O., Davis, M., Huchra, J., & Dressler, A., 1995, ApJ, in press
Saunders, W. *et al.* 1994, in preparation
Schechter, P. 1980, ApJ, 85, 801
Strauss, M. 1989, Ph.D thesis, UC Berkeley
Strauss, M., & Davis, M. 1988, in *Large-Scale Motions in the Universe*, ed. V. Rubin & G Coyne (Princeton University Press), p.255
Strauss, M., Huchra, J., Davis, M., Yahil, A., Fisher, K., & Tonry, J. 1992, ApJS, 83, 29
Strauss, M., & Yahil, A. 1994, unpublished
Strauss, M., & Willick, J., 1995, Physics Reports, in press
Therrien, C. W. 1992, *Discrete Random Signal and Statistical Signal Processing*, (New Jersey: Prentice-Hall)
Tonry, J. *et al.* 1994, in preparation
Vogeley, M. 1994, preprint
White, S., Navarro, J., Evrard, A., & Frenk, C., 1993, Nature, 366, 429
Willick, J. *et al.* 1994a, ApJ, to be submitted
Willick, J. *et al.* 1994b, ApJ, to be submitted
Yahil, A., Strauss, M., Davis, M., & Huchra, J. 1991, ApJ, 372, 380